\documentstyle[12pt,bezier]{article}
\begin{document}
% ***************    NEW COMMANDS   *******************
\def \inbar{\vrule height1.5ex width.4pt depth0pt}
\def \xC{\relax\hbox{\kern.25em$\inbar\kern-.3em{\rm C}$}}
\def \xR{\relax{\rm I\kern-.18em R}}
\newcommand{\xZ}{Z \hspace{-.08in}Z}
\newcommand{\xbe}{\begin{equation}}
\newcommand{\xee}{\end{equation}}
\newcommand{\xbea}{\begin{eqnarray}}
\newcommand{\xeea}{\end{eqnarray}}
\newcommand{\xnn}{\nonumber}
\newcommand{\xkt}{\rangle}
\newcommand{\xbr}{\langle}
\newcommand{\cun}{{\mbox{\tiny${\cal N}$}}}
\title{Quantum Adiabatic Approximation,  Quantum Action,~and~Berry's~Phase}
\author{Ali Mostafazadeh\thanks{E-mail: alimos@phys.ualberta.ca}\\ \\
Theoretical Physics Institute, University of Alberta, \\
Edmonton, Alberta,  Canada T6G 2J1.}
\date{June 1996}
\maketitle

\begin{abstract}
An alternative interpretation of the quantum adiabatic approximation is
presented.  This interpretation is based on the ideas originally advocated
by David Bohm in his quest for establishing a hidden variable alternative
to quantum mechanics. It indicates that the validity of the quantum adiabatic
approximation is a sufficient condition for the separability of the quantum
action function in the time variable. The implications of this interpretation
for Berry's adiabatic phase and its semi-classical limit are also discussed. 
\end{abstract}
\vspace{1cm}

Probably one of the best recognized applications of the quantum 
adiabatic approximation \cite{messiah,p16} is in Berry's derivation of the
adiabatic geometrical phases \cite{berry1984}.  Following Berry's, by now,
classical article on the adiabatic geometric phase \cite{berry1984}, Hannay
proposed a classical analogue of Berry's phase \cite{hannay} and Berry
\cite{berry1985} and Anandan \cite{anandan1988} explored the semiclassical
limit of Berry's phase and the classical analogue of the general, non-adiabatic
geometric phase \cite{aa}, respectively. The purpose of this note is to provide
an alternative interpretation of the quantum adiabatic approximation which
yields a natural approach to study the semi-classical limit of this approximation
and consequently Berry's phase. 

Consider the parameter-dependent quantum Hamiltonian:
	\xbe
	\hat H[R]=H(\hat p,\hat x;R):=\frac{1}{2}\, g^{ab}(R)
	[\hat p_a-A_a(\hat x;R)][\hat p_b-A_b(\hat x;R)]+V(\hat x;R)\;,
	\label{qu-ha}
	\xee
where $\hat x=(\hat x^1,\cdots,\hat x^\cun)$ and
$\hat p=(\hat p^1,\cdots,\hat p^\cun)$ denote the position and momentum
operators,  $R=(R^1,\cdots,R^m)$ are the coordinates of a parameter
space $M$, $g^{ab}(R)$ are the entries of a positive-definite symmetric
invertible $R$-dependent matrix, and $A_a$ and $V$ are arbitrary
vector and scalar potentials.  Furthermore, let $\{|n;R\xkt\}$ be a
complete orthonormal set of eigenvectors of $\hat H[R]$:
	\xbe
	\hat H[R]|n;R\xkt=E_n[R]|n;R\xkt\;.
	\label{eg-va-eq}
	\xee
Then a curve $C:[0,T]\to M$ defines a time-dependent Hamiltonian
according to: $\hat H(t):=\hat H[R_C(t)]$, where $R_C(t)$ are coordinates
of $C(t)$. In this case the energy eigenvectors and eigenvalues also
become time-dependent: $|n;t\xkt:=|n;R_C(t)\xkt$, $E_n(t):=E_n[R_C(t)]$.

The dynamics of the corresponding quantum system is governed
by the Schr\"odinger equation:	
	\xbe
	i\hbar\:\frac{d}{dt}\,|\psi(t)\xkt=\hat H(t)|\psi(t)\xkt\;,
	~~~~~~|\psi(0)\xkt=|\psi_0\xkt\;.
	\label{sch-eq}
	\xee
The quantum adiabatic approximation \cite{messiah} states that if the
Hamiltonian depends on time adiabatically \cite{p16}, then an eigenstate 
of $\hat H(0)$ evolves into an eigenstate of $\hat H(t)$. In particular if
$E_n(t)$ is non-degenerate for all $t\in[0,T]$, then $|\psi_0\xkt=|n;t=0\xkt$
implies \cite{berry1984}:
	\xbea
	|\psi(t)\xkt&\approx&e^{i\alpha_n(t)}|n;t\xkt\;,~~~~~\alpha_n(t)\:=\:
	\delta_n(t)+\gamma_n(t)
	\;,
	\label{psi=}\\
	\delta_n(t)&:=&-\frac{1}{\hbar}\int_0^t E_n(t')dt'\;,~~~~~
	\gamma_n(t)\:=\:\int_0^t i\xbr n;t|\frac{d}{dt}|n;t\xkt\:=\:
	\int_{R_C(0)}^{R_C(t)} A_n[R]\;,
	\label{de-ga}\\
	A_n[R]&:=&i\xbr n;R|d|n;R\xkt\::=\:i\xbr n;R|\frac{\partial}{\partial R^j}
	|n;R\xkt\:dR^j\;.
	\label{co}
	\xeea
Here the symbol $\approx$ is used to emphasize that the corresponding
relation is an approximation. 
If the curve $C$ is closed, i.e., $C(T)=C(0)$ then $\hat H(t)$ is a periodic
Hamiltonian. In this case, the phase angles $\alpha_n(T)$, $\delta_n(T)$,
and $\gamma_n(T)$, are called the adiabatic total, dynamical, and
geometrical (Berry) phase angles.

In order to investigate the semi-classical (WKB) limit of the quantum 
adiabatic approximation, consider the position representation of the
quantum system. In this case, the state vectors $|\psi\xkt$ are represented
by wave functions $\xbr x|\psi\xkt=\psi(x)$. In this article, I shall only
consider the case where $\psi$ is a complex-valued square integrable
function of  $x\in \xR^\cun$. 

Following Bohm \cite{d-bohm}, one can write $\psi=\sqrt{\rho}
\exp[iS/\hbar]$, where $\rho$ and $S$ are real-valued functions
and $\rho$ is positive semi-definite. In terms of $\rho$ and $S$ the 
Schr\"odinger equation (\ref{sch-eq}) is written in the form:
	\xbea
	\partial_t S(x;t)+ H(x,\nabla S(x;t);t)+Q(x;t)&=&0\;,
	\label{1}\\
	\partial_t\rho(x;t)+ \nabla\cdot J(x;t)&=&0\;,
	\label{2}
	\xeea
where $H(x,\nabla S(x;t);t)$ is the classical Hamiltonian $H(x,p;t)$
evaluated at $p=\nabla S(x;t)$, $Q:=-\hbar^2 g^{ab}
(\partial_a\partial_b\sqrt{\rho})/(2\sqrt{\rho})$ is the quantum
potential, and $J$ is the probability current density whose components are
defined by $J^a:=\rho g^{ab}(\partial_b S-A_b)$. There are two well-known 
properties of Eqs.~(\ref{1}) and (\ref{2}).
Firstly, for $Q=0$, Eq.~(\ref{1}) becomes identical with the Hamilton-Jacobi
equation \cite{goldstein}. Therefore, in this case the phase angle $S=S(x;t)$
of the wave function $\psi(x;t)$ is nothing but the classical action
function. This justifies the name ``quantum action'' for the general
case where $Q$ does not vanish. In this general case, one can attempt
to solve  (\ref{1}) and (\ref{2}) by an iterative method in the first step of
which one approximates $Q$ by zero. This approximation is know as the
semi-classical or WKB approximation, \cite{d-bohm-qm,schiff}. Secondly,
Eq.~(\ref{2}) is a continuity equation corresponding to the conservation
of the probabilities. 

Similarly, one can express the eigenvalue equation (\ref{eg-va-eq})
in the position representation. Using the notation 
	\[
	\xbr x|n;R\xkt=:\psi_n(x;R)=:\sqrt{\rho_n(x;R)}\:e^{iS_n(x;R)/\hbar}\;,\]
	\[Q_n:=-\frac{\hbar^2\:g^{ab}(\partial_a\partial_b\sqrt{\rho_n})}{
	2\sqrt{\rho_n}}\:, ~~~~J^a_n:=\rho_n g^{ab}(\partial_b S_n-A_b)\,,\]
one has:
	\xbea
	H(x,\nabla S_n(x;R);R)+Q_n(x;R)&=& E_n[R]\;,
	\label{3}\\
	\nabla\cdot J_n(x;t)&=&0\;.
	\label{4}
	\xeea
Again in the semi-classical limit (also indicated by $\hbar\to 0$ in the
loop expansion of the path integral kernel \cite{bd}) $Q_n$ is neglected
\cite{d-bohm-qm,schiff}.

Next let us examine the statement of the quantum adiabatic approximation
in the position representation.  In view of (\ref{psi=}), one has:
	\xbe
	\rho(x;t)\approx\rho_n(x;R(t))\,,~~~~S(x;t)\approx S_n(x;R(t))+
	\hbar\:\alpha_n(t)\;.
	\label{5}
	\xee
In order to express $\alpha_n$ in terms of $\rho_n$ and $S_n$, one may naively
attempt to differentiate both sides of the second relation in (\ref{5}) and
use Eqs.~(\ref{1})-(\ref{5}) to simplify the result. This would lead to a
differential equation for $\alpha_n$ which implies the incorrect result:
 $\alpha_n=\delta_n$. The problem with this derivation is that  the
relations in (\ref{5}) are not  strict equations. They are only valid under
the conditions of the adiabatic approximation \cite{p16}. In other words, in
general there are other additive terms on the right hand sides of these relations
which  in the adiabatic approximation may be neglected. The time derivatives
of these additional terms however may not be negligible. Therefore one is allowed to
use the relations in (\ref{5}) as equalities as far as no time-differentiation
is directly performed.  For example, one has $\nabla S\approx\nabla S_n$, 
$Q\approx Q_n$ by virtue of which Eqs.~(\ref{1}) and (\ref{2}) take the form:
	\xbe
	\partial_t S(x;t)\approx-E_n(t)\;,~~~~~\partial_t\rho(x;t)\approx 0\;.
	\label{1'}
	\xee
Here use is also made of Eqs.~(\ref{3}) and (\ref{4}). These equations can be
integrated to yield: 
	\xbe
	S(x;t)\approx S_n(x;0)+f(t)\;,~~~~~\rho(x;t)\approx \rho_n(x;0)\;,
	\label{1''}
	\xee
where $f(t):=-\int_0^tE_n(t')dt'$. Eqs.~(\ref{1''}) indicate that for an 
adiabatic quantum system, the quantum action $S(x;t)$  with $S(x;0)$
corresponding to an initial energy eigenvector, is separable in the
time variable, whereas the quantum potential is time-independent.

Next consider the transition amplitude
	\xbe
	\xbr n;t|\psi(t)\xkt=:e^{i\alpha_n(t)}\;,
	\label{6}
	\xee
which may be viewed as the definition of $\alpha_n(t)$. Expressing (\ref{6})
in the position representation and differentiating the result with respect to time
one obtains:
	\xbe
	\frac{d}{dt}\:\alpha_n(t)=-i\int dx^\cun e^{\frac{i}{\hbar}(S-S_n-\hbar\alpha_n)}
	\left[ \frac{1}{2}(\sqrt{\frac{\rho}{\rho_n}}\partial_t\rho_n+
	\sqrt{\frac{\rho_n}{\rho}}\partial_t\rho)+\frac{i}{\hbar}\sqrt{\rho_n\rho}\:(
	\partial_t S-\partial_t S_n)\right]\;.
	\label{7}
	\xee
One can now use (\ref{1}), (\ref{1'}), (\ref{5}) and $\int dx^\cun \rho=
\int dx^\cun \rho_n=1$ to simplify the integrand in (\ref{7}). This yields:
	\xbe
	\frac{d}{dt}\: \alpha_n(t)=-\frac{1}{\hbar}\:E_n(t)-
	\frac{1}{\hbar}\int dx^\cun \: \rho_n(x;R(t))\:\partial_t S_n(x;R(t))\;,
	\label{8}
	\xee
which up on integration reproduces Berry's result (\ref{psi=}) with:
	\xbe
	A_n[R]=-\frac{1}{\hbar}\int dx^\cun \:\rho_n(x;R)\:\frac{\partial}{\partial R^j}\,
	S_n(x;R)\:dR^j\;.
	\label{9}
	\xee
An obvious implication of this equation is that if the energy eigenfunctions
are real, such as in the case of simple Harmonic oscillator, then the
geometric phase is identically zero. 

In the semi-classical approximation, one has $Q(x;t)=Q_n(x;t)=0$ and
the quantum action becomes identified with the classical action function.
Therefore the requirement of the adiabaticity of the Hamiltonian leads
to the separability of the classical action in the time variable. In particular,
if the energy eigenvalues are time-independent, then $S(x;t)\approx
S_n(x;0)-E_n\,t$ which is exactly valid when the (classical) Hamiltonian
is time-independent. In this case the $S_n(x;0)$ is called Hamilton's 
characteristic function \cite{goldstein}. 

Another implication of the adiabatic approximation is that the validity
of the semi-classical approximation can be directly checked by computing
the initial quantum potential $Q(x;0)$. If $Q(x;0)$ is negligibly small then
by virtue of  $\rho(x,t)\approx \rho(x;0)$ it remains small throughout an
adiabatic evolution, and the semi-classical approximation is valid.

\end{document}